\documentstyle[12pt,epsf,epsfig,wrapfig]{article}
\begin{document}

\begin{center}
{\bfseries DIRECTED FLOW AND ROTATION OF TRANSIENT MATTER }

\vskip 5mm \underline{S.M. Troshin}$^\dag$ and N.E. Tyurin

\vskip 5mm
{\small
 {\it
Institute for High Energy Physics, Protvino, 142281 Russia
}\\
$^\dag${\it E-mail: Sergey.Troshin@ihep.ru }}
\end{center}

\vskip 5mm
\begin{abstract}
Directed flow $v_1$ treated as  an effect
of the transient matter rotation in hadronic and nuclei reactions.
\end{abstract}

\vskip 8mm
Multiparticle production in hadron and nucleus collisions and
corresponding observables  provide  a clue to the mechanisms of
confinement and hadronization. Discovery of the deconfined state
of matter has been announced  by the four major experiments at RHIC
\cite{rhic}. Despite the highest values of energy and density have
been reached, a genuine quark-gluon plasma QGP (gas of the free
current quarks and gluons) was not found \footnote{It is  to be
noted here that confinement due to causality principle might exclude the
 very existence of QGP defined
that way \cite{mrowch}.}. The deconfined state
reveals the properties of the perfect liquid, being strongly
interacting collective state and therefore it was labelled as sQGP
\cite{denteria}.  The nature of the new form of
matter  is not known. The importance of the experimental discoveries
at RHIC
is that the matter remains  strongly correlated
and reveals high degree of the coherence when it is
 well beyond the critical values of density and
temperature.
In this report we would like to stress that the behavior of collective observables
 in hadronic and nuclear
reactions could have some similarities. Among several experimental probes  of collective dynamics in
$AA$ interactions \cite{voloshin,molnar} are the momentum anisotropies
$v_n$  defined
by means of the Fourier expansion of the transverse momentum spectrum
over the momentum azimuthal angle $\phi$. With  measurements of these observables
one can obtain a valuable information on  the early stages of reactions and observe
signals of QGP formation \cite{s1,s2,s3,s4,s5,s6,s7,s8,s9}. We discuss the role of the coherent rotation
of the transient matter in hadron and nuclei collisions and  the directed
flow dependence.
Hypothesis on connection
  of the strongly interacting transient matter rotation with the directed flow generation is the
  main point of this report.

 We  consider non-central hadron collisions
and apply notions acquired from heavy-ion studies.
 In particular, we  amend the model \cite{csn} developed for hadron interactions
 (based on the chiral quark
model ideas)  and consider the effect of collective rotation of a quark matter in the overlap region.
The
determination of the reaction plane in the non-central hadronic collisions \cite{polvol} could be
 experimentally realizable   with the utilization of the standard procedure\cite{poskanz}.
 Geometrical picture of hadron collision at non-zero impact parameters \cite{csn}
implies that the generated massive
virtual  quarks in overlap region  (due to shock-wave type of interaction of the
condensate clouds\footnote{This mechanism is similar to the shock-wave production process proposed by
Heisenberg\cite{sh}}) carry large orbital angular momentum
at high energies. The total orbital angular
momentum  can be estimated
as follows
\begin{equation}\label{l}
 L(s,b) \simeq \alpha b \frac{\sqrt{s}}{2}D_C(b).
\end{equation}
The parameter $\alpha$ is related to the fraction of the initial energy carried by the condensate
clouds which goes to rotation of the quark system and
\[
D_C(b)\equiv D_c^{h_1}\otimes D_c^{h_1},
\]
where function $D^h_c$
describes peripheral condensate distribution inside the hadron $h$, and $b$ is
an impact parameter of the colliding hadrons.
The overlap region, which is described by the function $D_C(b)$,
has an ellipsoidal form similar to the overlap region in the nucleus collisions.
It should be noted that $L\to 0$ at $b\to\infty$ and $L=0$ at $b=0$.
Similar impact parameter dependence with maximum at the impact parameter values around 1 fm
the directed flow $v_1$ has.

Due to strong interaction
between quarks the orbital angular momentum $L$ leads to  coherent rotation
of the quark-pion liquid located in the overlap region as a whole  in the
$xz$-plane
since strong correlations between particles  are presented there. It should
be noted that for the given value of the orbital angular momentum $L$
 kinetic energy has a minimal value if all parts of liquid rotates with the same angular velocity.
  We  assume therefore
that the different parts of the quark-pion liquid in the overlap
region indeed have the same angular velocity $\omega$. It has
grounds also in the perfect, non-viscous, character of the liquid
revealed at RHIC. Such coherent rotation is absent in the parton
picture used in \cite{wang}, where  finite transverse gradient of parton
longitudinal momentum  is a driving force of the orbital angular
momentum conversion to the global system polarization through
spin-orbital coupling. The polarization
not yet been detected experimentally \cite{starpol}.

The generation time of the transient state $\Delta t_{tsg}$ obey to the inequality
$\Delta t_{tsg}\ll \Delta t_{int}$,
where $\Delta t_{int}$ is the total interaction time.
The assumed particle production mechanism at moderate transverse
momenta is
an excitation of  a part of the rotating transient state of  massive constituent
quarks (interacting by pion exchanges) by the one of the valence constituent quarks
 with  subsequent hadronization
of the quark-pion liquid droplets.
Due to the fact that the transient matter is strongly interacting, the excited parts
should be located closely  to the periphery  of the rotating transient state otherwise absorption
 would not allow
to quarks and pions to leave the region (quenching). The mechanism is sensitive
 to the particular
rotation direction and the directed flow should have  opposite signs for the particles
in the fragmentation regions of the projectile and target respectively.
It is evident that the effect of  rotation (shift
in  $p_x$ value ) is
most significant in the peripheral part of the rotating quark-pion liquid
and is to be weaker in the less peripheral regions (rotation with the same angular velocity $\omega$),
i.e. the directed flow $v_1$ (averaged over all transverse
momenta)  directly depends on the distance to the center of the rotating matter or on the depth $\Delta l$
where the excitation of the rotating quark-pion liquid takes place.
In its turn, the length   $\Delta l$ should be proportional to the energy loss of constituent
valence quark in the medium (quark-pion liquid) prior an excitation occurs, i.e. before constituent
quark would deposit its energy into the energy of the excited quarks (those quarks lead to the production
of the secondary particles)
\begin{equation}\label{ener}
\Delta l\sim \Delta E.
\end{equation}
Proportionality of the energy loss due to elastic rescattering and $\Delta l$
 is a consequence of the liquid
nature of the transient state which has  fixed interparticle distances. Energy
loss $\Delta E$
should  (in a rough approximation) be proportional to the
difference between the rapidities  of the final particle and  the projectile.
Thus, the observable $v_1$, which magnitude is determined by the shift of transverse momentum
due to rotation and depend therefore on the value of $\Delta l$ in this mechanism,
 would depend in its turn on  the rapidity difference
 $y-y_{beam}$ and not on the incident energy. The mechanism
 therefore can provide a qualitative explanation of the incident-energy scaling
 of $v_1$ observed at RHIC \cite{v1exp}.  Evidently,
  the directed flow $|v_1|$ decreases
  when the absolute value of the above difference increases, i.e. $|v_1|$ increases at
  fixed energy and increasing rapidity of final particle and it decreases at fixed rapidity
  of final particle and increasing  beam energy.

  An important assumption
  based on the RHIC  data is the strongly interacting nature of the transient matter, namely,
  it was supposed that valence constituent quark excites quark-pion liquid
   in the closest  hemisphere to the entry point.

The magnitude of $|v_1|$ is to be proportional to inverse depth length $\Delta l^{-1}$ which is determined
  by elastic quark scattering cross-section $\sigma$ and quark pion liquid density $n$, i.e.
$ \Delta l\sim 1/\sigma n$
and therefore averaged value of $v_1$ should be proportional to the particle density of
the transient state
$\langle |v_1|\rangle \sim \sigma n$.
   This estimate shows that the magnitude of the directed flow could provide information
   on the properties of the transient state.
  The centrality dependence of $v_1$ should   be decreasing towards  high and lower centralities.
  Decrease toward high centralities is evident, no overlap of hadrons or nuclei should be at high enough
  impact parameters. Decrease of $v_1$ toward lower centralities is specific prediction of the proposed
  mechanism based on rotation
  since central collisions with smaller impact parameters would lead to slower rotation or its complete
   absence in the head-on collisions.

If the proposed mechanism of the directed flow generation is realized,
 vanishing directed flow can serve as a  signal of a genuine quark-gluon plasma
(gas of free quarks and gluons) formation. Then the orbital momentum,  could be converted
e.g. into the global polarization at the partonic
level  and detected experimentally  measuring hyperon or photon polarizations \cite{wang}.

It
would be interesting to perform studies of transient matter at the
LHC not only in heavy ion collisions, but also in $pp$--collisions,
and to find possible existence or absence of the rotation effects through
the directed flow and polarization measurements.
Collective rotation
  should also contribute to the elliptic flow. However, since the regularities already found experimentally for
  $v_1$ and $v_2$ in nuclei interactions imply  different dynamical origin for these flows, we should conclude
  that the rotation does not provide a significant contribution to elliptic flow.
\section*{Acknowledgement}
We are grateful to J. Dunlop and O. Teryaev for the interesting discussions. One of the authors (S.T.) is also
grateful to the Organizing Committee of the XII Workshop on High Energy Spin Physics and especially to its Chairman
A.V. Efremov  for the support and warm hospitality in Dubna.


\begin{thebibliography}{99}
\bibitem{rhic}
 Quark Gluon Plasma. New Discoveries at RHIC: A Case of Strongly Interacting Quark Gluon Plasma.
 Proceedings, RBRC Workshop, Brookhaven, Upton, USA, May 14-15, 2004:
D. Rischke, G. Levin, eds; 2005, 169pp; J. Adams et al. (STAR Collaboration)
Nucl. Phys. A 757, 102 (2005);
K. Adcox et al (PHENIX Collaboration), Nucl. Phys. A 757,  184 (2005).
\bibitem{mrowch}
D. Mi\'skoviec, [arXiv: 0707.0923].
\bibitem{denteria}
D. d'Enterria, [arXiv: nucl-ex/0611012].
\bibitem{voloshin}
J.-Y. Ollitrault,
Phys. Rev. D46,  229 (1992); recent review can be found in
S.A. Voloshin, Nucl. Phys. A715,  379 (2003).
\bibitem{molnar}
D. Molnar, [arXiv: nucl-th/0408044].
\bibitem{s1}
 J.~Hofmann, H.~St\"ocker, U.~W.~Heinz, W.~Scheid, W.~Greiner,
Phys.\ Rev.\ Lett.\  36,  88, (1976).
\bibitem{s2}
 H.~St\"ocker, J.~A.~Maruhn, W.~Greiner,
 Phys.\ Lett.\  B 81, 303,  (1979).
\bibitem{s3}
 H.~St\"ocker, J.~A.~Maruhn, W.~Greiner,
 Phys.\ Rev.\ Lett.\  { 44}, 725  (1980).
\bibitem{s4}
 H.~St\"ocker, W.~Greiner,
 Phys.\ Rept.\   137,  277, (1986).
\bibitem{s5}
 H.~Sorge,
 Phys.\ Rev.\ Lett.\  82, 2048  (1999)
 [arXiv:nucl-th/9812057].
\bibitem{s6}
 M.~Bleicher, H.~St\"ocker,
 Phys.\ Lett.\  B  526,  309 (2002)
 [arXiv:hep-ph/0006147].
\bibitem{s7}
 H.~St\"ocker,
 Nucl.\ Phys.\  A  750,  121 (2005)
 [arXiv:nucl-th/0406018].
\bibitem{s8}
 X.~l.~Zhu, M.~Bleicher, H.~St\"ocker,
 J.\ Phys.\ G 32, 2181  (2006)
 [arXiv:nucl-th/0601049].
\bibitem{s9}
 H.~Petersen, Q.~Li, X.~Zhu, M.~Bleicher,
 Phys.\ Rev.\  C  74,  064908 (2006)
 [arXiv:hep-ph/0608189].
\bibitem{csn}
S. M. Troshin, N. E.Tyurin, Phys. Rev. D  49,  4427 (1994).
\bibitem{polvol}
S.A. Voloshin, [arXiv: nucl-th/0410089].
\bibitem{poskanz}
S.A. Voloshin, A.M. Poskanzer, Phys. Lett. B 474,  27 (2000).
\bibitem{sh}
W. Heisenberg, Zeit. Phys. 133, 65  (1952).
\bibitem{starpol}
B.I. Abelev et al, (STAR Collaboration), [arXiv:0705.1691].
\bibitem{wang}
Z.-T. Liang, X.-N. Wang,
Phys. Rev. Lett. 94,  102301 (2005), ibid. 96,  039901 (2006)\\
A. Ipp, A. Di Piazza, J. Evers, C.H. Keitel, [arXiv:0710.5700].
\bibitem{v1exp}
G. Wang (for the STAR Collaboration), arXiv: nucl-ex/0701045;//
B.B. Back et al., (PHOBOS Collaboration), Phys. Rev. Lett., 97,  012301 (2006);\\
A.H. Tang (for the STAR collaboration), J. Phys. G: Nucl. Part. Phys. 31,  S35 (2005);\\
J. Adams et al. (STAR Collaboration),  Phys. Rev. C 73,  034903 (2006).\\
\end{thebibliography}
\end{document}